\documentclass[aps,prl,superscriptaddress,showpacs,twocolumn,floatfix]{revtex4}
\usepackage{amsmath}
\usepackage{amssymb}
\usepackage{graphicx}
\usepackage{epsfig}

\newcommand {\be}{\begin{equation}}
\newcommand {\ee}{\end{equation}}
\newcommand {\bea}{\begin{eqnarray}}
\newcommand {\eea}{\end{eqnarray}}
\newcommand {\sech}{\text{sech}}

\begin{document}

\title{Friction and diffusion of matter-wave bright solitons}
\author{Subhasis Sinha}
\affiliation{Max Planck Institute for the Physics
of Complex Systems, N{\"o}thnitzer Stra{\ss}e 38, 01187 Dresden, Germany}

\author{Alexander Yu.~Cherny}
\affiliation{Max Planck Institute for the Physics
of Complex Systems, N{\"o}thnitzer Stra{\ss}e 38, 01187 Dresden, Germany}
\affiliation{Bogoliubov Laboratory of Theoretical Physics, Joint
Institute for Nuclear Research, 141980, Dubna, 
Russia}

\author{Dmitry Kovrizhin}
\affiliation{Max Planck Institute for the Physics
of Complex Systems, N{\"o}thnitzer Stra{\ss}e 38, 01187 Dresden, Germany}
\affiliation{RRC Kurchatov Institute, Kurchatov Sq.~1, 123182 Moscow, Russia}

\author{Joachim Brand}
\affiliation{Max Planck Institute for the Physics
of Complex Systems, N{\"o}thnitzer Stra{\ss}e 38, 01187 Dresden, Germany}

\begin{abstract}
We consider the motion of a matter-wave bright soliton under the
influence of a cloud of thermal particles. In the ideal
one-dimensional system, the scattering process of the quasiparticles
with the soliton is reflectionless, however, the quasiparticles acquire
a phase shift. In the realistic system of a Bose-Einstein condensate
confined in a tight waveguide trap, the transverse degrees of freedom
generate an extra but small nonlinearity in the system which gives
rise to finite reflection and leads to dissipative motion of the
soliton. We calculate the velocity and temperature-dependent
frictional force and diffusion coefficient of a matter wave bright
soliton immersed in a thermal cloud.
\end{abstract}

\maketitle

Solitons are localized waves that propagate without spreading and
attenuation. They appear from classical systems like ocean waves to
optics and quantum systems like
Bose-Einstein condensates (BEC)
of atomic gases. A BEC of a dilute atomic gas with attractive two-body
interactions in three dimensions (3D) is unstable and collapses
\cite{gerton00}.  In one dimension (1D), however, a BEC with
attractive interaction is stable against collapse and forms a
self-bound particle-like object known as a bright soliton.  Recently,
bright solitons of Bose-condensed $^{7}$Li atoms were observed in
quasi-1D waveguide traps at Rice University \cite{Strecker2002a} and
at ENS in Paris \cite{Khaykovich2002a}.

One of the most important features of solitons is the
dissipationless motion over long distances. Because
of this property, optical solitons have important applications in
transatlantic fiber-optic communication systems \cite{hasegawa02}.
In this Letter we discuss how dissipative effects in the motion of a
soliton in a thermal cloud 
can arise due to the 3D 
nature of the BEC in a tight waveguide.
Although the dynamics is strictly one-dimensional, the transverse
extent of the mean field generates extra nonlinear terms in the
effective 1D equation, as first discussed in Ref.~\cite{Muryshev2002a}.
This formalism can also be applied to
other systems described by a cubic nonlinear Schr{\"o}dinger
equation perturbed by a quintic nonlinear term. An overview of related
work in nonlinear optics can be found in Ref.~\cite{kivshar03:book}.

In this Letter we consider the scattering of quasiparticles by the
quasi-1D matter-wave soliton \cite{Wynveen2000a}
similar to linear waves scattering on breathers \cite{flach:084101}.
Muryshev {\it et al.} considered the interaction of quasiparticles
with dark BEC solitons and conjectured that these lead to acceleration and
eventually disintegration of the soliton in a thermal
environment~\cite{Muryshev2002a}.
Following  a similar line of arguments, we show that the quasiparticles
scattering on a bright soliton have a finite probability of reflection
only due to the extra nonlinearity, which finally gives rise to dissipative
effects.
The bright soliton
experiences friction and diffusive motion in a thermal cloud but
maintains its integrity in
contrast to dark solitons which disintegrate.

A BEC in a waveguide with a harmonic transverse confinement is well
described by the Gross-Pitaevskii (GP) equation:
\begin{equation} \label{GP3}
i\hbar\frac{\partial \psi}{\partial t} = \Big[-\frac{\hbar^{2}}{2 m}
\nabla^2  + \frac{1}{2}m
\omega^{2} \rho^{2} + \frac{4 \pi \hbar^2 a}{m} |\psi|^2\Big]
\psi
\end{equation}
where $\psi$ is the macroscopic wavefunction of the condensate,
$\omega$ is the frequency of transverse trapping potential, and
$a$ is the 3D scattering length. In the quasi-1D limit
the effective dynamics of the system takes place along the free
axis (x-axis) without exciting the transverse modes. The quasi-1D limit
can be achieved when the mean field interaction is smaller than the
radial excitation frequency, ${4 \pi \hbar^2 |a|} |\psi|^2/ {m}  < \hbar
\omega$.
Aiming at an adiabatic separation of slow longitudinal and fast
transverse motion we
can write the full 3D wave function  assuming cylindrical symmetry as,
$\psi(\vec{r},t) = \phi(x,t)\chi(\rho,x,t)$.
Here, $\phi$ is the 1D (longitudinal) wavefunction and $\chi$ is the
radial wavefunction with the normalisation convention $\int |\chi|^2
2\pi \rho\,d\rho = 1$ and $\int |\phi|^2 dx = N$, where $N$ is
the number of bosons in the system.
In the adiabatic or Born-Oppenheimer approximation we now assume that
the radial wavefunction $\chi$ depends only weakly on the slow
variables $x$ and $t$ and their derivatives of $\chi$ can be
neglected.  For this assumption to be correct, the time scale of
dynamics should be longer than the inverse of the tranverse frequency
$\omega$ and the relevant length scale
should be significantly
larger than the transverse oscillator length $l = \sqrt{\hbar/(m
\omega)}$.
After substituting the ansatz for
the wavefunction into Eq.~(\ref{GP3}) and neglecting the derivatives
of $\chi$ with respect to $x$ and $t$, we obtain the following
adiabatically decoupled equations for the longitudinal and the
transverse wavefunctions:
\begin{align}\label{eqn:long}
i\hbar \partial_t \phi &=  -\frac{\hbar^{2}}{2 m} \partial_{x}^{2}\phi +
\tilde{\mu}\phi ,\\
\label{eqn:transverse}
 \Big[-\frac{\hbar^{2}}{2 m}\nabla^2_{\rho}
+& \frac{1}{2}m
\omega^{2} \rho^{2}
+ \frac{4 \pi \hbar^2
a}{m} n
|\chi|^2\Big] \chi \hfill
= \tilde{\mu} \chi ,
\end{align}
where we have introduced the transverse chemical
potential $\tilde{\mu}$, which has to be found from the ground
state solution of Eq.~(\ref{eqn:transverse}) as a function of the linear
density $n(x,t) = |\phi(x,t)|^2$. A simple scaling argument shows that
$\tilde{\mu} = \hbar\omega f(a n)$, where
$f(\cdot)$ is a
dimensionless function, which has been computed numerically in
Ref.~\cite{Berge2000a}. Physical solutions of Eq.~(\ref{eqn:transverse})
are found only if $- a n <0.47$ \cite{weinstein83,Berge2000a}, otherwise
transverse collapse occurs \cite{carr:040401}. In the
following we will be interested in the quasi-1D regime of small $a n$
and expand $\tilde{\mu}(a n)$ in a power series.

In the quasi-1D limit, when $|a| n \ll 0.47$,
the radial wavefunction $\chi$
will be close to the ground state of the 2D harmonic
oscillator with a Gaussian profile.
We can expand $\chi$ in terms of the radial eigenmodes
$\varphi_\nu(\rho)$, $\chi(\rho, x) = \varphi_{0}(\rho) + \sum_{\nu}
C_{\nu}(x) \varphi_{\nu}(\rho)$.
The coefficients $C_\nu$ are small and can be calculated perturbatively.
The transverse chemical potential $\tilde{\mu}$ can be obtained by using
second order perturbation theory:
\begin{equation} \label{eqn:nlexpansion}
\tilde{\mu} = \hbar \omega + g n - g_2 n^2 +
\ldots ,
\end{equation}
where $g = 2 a \hbar \omega$ and $g_2 = 24 \ln(4/3) a^2 \hbar
\omega$.
A correction to the 1D coupling constant $g$ beyond the GP
approach presented here has been found in
Ref.~\cite{Olshanii1998a}. The constant $g_2$ was calculated first in
Ref.~\cite{Muryshev2002a} and corrections beyond GP can be obtained
by the self-consistent Hartree Fock Bogoliubov approach of
Ref.~\cite{cherny:043622}.
We obtain the following effective equation
describing the condensate in the quasi-1D limit:
\begin{equation} \label{1d}
[-({\hbar^{2}}/{2 m}) \partial_{x}^{2} + g |\phi|^2 - g_2 |\phi|^4] \phi
= \mu \phi .
\end{equation}
This is a nonlinear Schr\"odinger equation with a cubic and a quintic
nonlinearity, as used before in Ref.~\cite{Muryshev2002a}. The possibility of
collapse is inherent in this equation as the quintic nonlinear term is
attractive. An estimate from the 3D GP equation (\ref{GP3}) gives
stability of a single soliton solution if $N |a|/ {l} < 0.627$ is
fulfilled \cite{carr02}.

Without the extra nonlinearity associated with $g_2$, Eq.~(\ref{1d})
is integrable. For attractive interactions at
$a<0$, the bosons form a self-bound particle-like
state known as a {\em bright soliton} with the wavefunction $\phi(x) =
\sqrt{N/2b}\, \sech(x/b)$ and the chemical
potential $\mu = -\hbar^2/2 m b^2$,
where $b = l^2/(N|a|)$. We notice that for a weak soliton parameter
$N|a|/l \lesssim 1$, the system becomes quasi-one-dimensional ($b \gtrsim l$).

A soliton can be considered as a macroscopic particle of mass $m N$,
moving in the bath of thermal excitations. Dissipative motion of the soliton
arises due to the scattering of thermal atoms.
Here we consider the interaction of thermally excited particles with the
soliton within the Bogoliubov formalism \cite{Wynveen2000a},
\begin{equation} \label{eqn:bog}
[H_0 + H_1]\psi = \epsilon(k) \psi
\end{equation}
where, $\psi = (u, v)$ is a two component vector of particle ($u$) and hole
($v$) amplitudes, and $\epsilon$ is the quasiparticle energy.
The unperturbed Hamiltonian $H_0$ and the perturbation $H_1$ are
given by
\begin{eqnarray}
\hat{H}_{0} & = & \left( \begin{array}{cc}
-\frac{\hbar^2}{2 m}\partial_{x}^{2}
- \mu & 0\\ 0 &
+\frac{\hbar^2}{2 m}\partial_{x}^{2} + \mu
\end{array} \right)\\
\hat{H}_{1} & = & \left( \begin{array}{cc} V_{1}(x)
& V_{2}(x)\\ -V^{*}_{2}(x) & -V_{1}(x)
\end{array} \right),
\end{eqnarray}
where $ V_{1} = 2g|\phi|^2 - 3 g_{2} |\phi|^{4}$ and
$V_{2} = g \phi^{2} - 2 g_{2}|\phi|^{2}\phi^{2}$.
The scattering states have energy $\epsilon(k) = \frac{\hbar^2 k^2}{2 m} +
|\mu|$.
In one dimension, neglecting
 the extra nonlinearity ($g_2 =0$), we
obtain
the exact solution of the scattering states:
\begin{align}\label{eqn:exactu}
u_{k} \! =& A(k)\, [kb + i\tanh(x/b)]^{2}e^{i k x}\\\label{eqn:exactv}
v_{k} \!= & A(k)\, \sech^{2}(x/b) e^{i k x} ,
\end{align}
where $A(k)= 1/(k^2 b^2 -1)$ is a normalisation constant.
The transmittance is given by
\begin{equation}
t = (k b + i)^{2}/(kb - i)^{2},
\end{equation}
and the transmission probability is $|t|^{2} = 1$.
Hence, the quasiparticles scatter without reflection on the soliton but only
aquire a phase shift and a time delay in the scattering process.
Reflectionless scattering on a soliton in the integrable nonlinear
Schr\"odinger equation (\ref{1d}) (with $g_2=0$) is a well-known result of
mathematical soliton theory and
is also found in an exact
solution of the quantum many-body model in the limit of large particle
number \cite{mcguire:622}.  In the quasi-1D limit, the soliton thus
becomes transparent
and exhibits dissipationless motion in a
thermal cloud.

\begin{figure}[htb]
\begin{center}
\epsfig{file=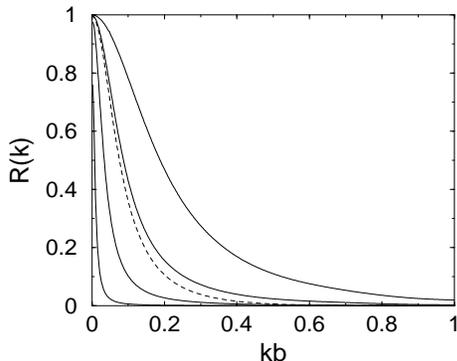,
height=5cm,width=6cm}
\end{center}
\hskip -2cm
\vspace*{-1.2cm}
\caption{Reflection coefficient of a soliton as a function of
momentum, for $N|a|/l =$ 0.1, 0.2, 0.3, 0.4 (from smaller to larger
values of reflection coefficient). The dotted line shows the
analytical estimate
(\ref{eqn:R}) for $N|a|/l = 0.2$.
\label{fig:1}}
\end{figure}

Now we consider the scattering problem of quasiparticles in the presence
of an extra nonlinearity that breaks the integrability.
Assuming that the coupling constant $g_2$ of the extra nonlinear term
is small, we
can solve the scattering problem using Green's function techniques.
In order to solve Eq.~(\ref{eqn:bog}) for the particle amplitude $u$ we
construct the Green's function for the $u$ component of $H_0$
satisfying
\begin{equation}
  [-({\hbar^2}/{2 m})\partial_{x}^{2} - \mu  - \epsilon]\; G_{1}(x-x')
  = -\delta(x-x') ,
\end{equation}
which is given
by $G_{1}(x -x') = (m/\hbar^{2} k) \sin(k |x - x'|)$.
Since the potential is symmetric, 
the scattering states
can be constructed with even or odd symmetry. The Lippmann-Schwinger
equation for the particle channel can be written as
\begin{eqnarray} \label{ls}
u_{e/o} & = & u_{e/o}^{0} + \int G_{1}(x -x')V_{1}(x')u_{e/o}(x')\,dx'
\nonumber\\ & & + \int
G_{1}(x -x')V_{2}(x')v_{e/o}(x')\,dx' ,
\end{eqnarray}
where $u_{e/o}$ denotes even [odd] wave functions of the particle states
and $u_{e}^{0} = \cos(kx)$, [$u_{o}^{0} =\sin(kx)$].
The most general wave function can be constructed from even and odd
eigenstates: $u_{k} = A u_{k}^{e} + B u_{k}^{o}$.
Asymptotically this wave function becomes $\lim_{x \to -\infty} u_{k}
=  e^{i k x} + r e^{-i k x}$ and   $\lim_{x \to \infty} u_{k}  =  t
e^{i k x}$
where $|t|^2$ and $R = |r|^2$ are the transmission and the reflection
coefficient,
respectively.
We obtain $R(k)$ by solving Eqs.~(\ref{1d},
\ref{eqn:bog}) 
numerically and matching with the asymptotic solutions,
see Fig.~\ref{fig:1}.

An analytical estimate of the reflection
coefficient can be obtained from Eq.~(\ref{ls}) by approximating $\phi =
\sqrt{N/2b}\,\sech(x/b)$ and $u_{e/o}$ and $v_{e/o}$ with the properly
symmetrised solutions (\ref{eqn:exactu}, \ref{eqn:exactv}). This
approximation becomes exact for $g_2=0$ and relies on $g_2$ being a
small parameter. The
reflection coefficient is given by $R=|r|^2$  and
\begin{equation} \label{eqn:R}
r(k)=-i\frac{I_{+}+I_{-}}{(I_{+}-i)(I_{-}+i)},
\end{equation}
where the terms $I_{+}$ and $I_{-}$ are given by
\begin{eqnarray}
I_{\pm} = \mp 2 A(k)\left[kb +6\ln(4/3)\frac{N^2|a|^2}{l^2}
  \frac{Q_{\pm}(kb)}{kb}\right]\nonumber 
\end{eqnarray}
with $Q_{\pm}(x) = 1/3+ x^{2}\pm (1 + x^2)^2\pi x/[3 \sinh(\pi x)]$.
By using the Lippmann-Schwinger formalism instead of a simple Born
approximation we obtain the correct
limiting behaviour for small $k$ where $R\to 1$.
Total reflection is expected whenever the
special resonant conditions leading to reflectionless scattering at $g_2=0$
are broken,
as $k\to 0$ implies a vanishing group velocity
$\partial \epsilon/\partial (\hbar k) = \hbar k/m$.
This case is very
different from phonons scattering on a perturbed dark soliton, which
becomes transparent for
small $k$
as found in
Ref.~\cite{Muryshev2002a}.
The approximation (\ref{eqn:R}) reproduces the qualitative features
but slightly overestimates the exact values of $R$ as seen in
Fig.~\ref{fig:1}.

The reflection coefficient $R$ is a function of
dimensionless momentum $kb$ and the soliton parameter $N|a|/l$. In
the dissipative dynamics of a macroscopic object like a soliton, the
microscopic parameter $N|a|/l$ enters through the reflection
coefficient of the quasiparticles.  Once we know the interaction of
particles with a soliton from the microscopic theory, we can describe
its motion in the bath of thermal particles at a given temperature.  A
bright soliton is a mesoscopic object with mass $mN$, and its dynamics
is governed by classical motion.
Therefore, we can define a phase space distribution function of
soliton's center of mass coordinate $f(p,q,t)$. When the soliton
follows the classical trajectory then the distribution function takes
a simple form $f(p,q,t) = \delta(p - p(t))\delta(q - q(t))$, where
$p(t), q(t)$ are classical phase space trajectories. In the
presence of a bath of thermal atoms, the atoms impart a momentum to
the soliton in the scattering process. While the soliton is at rest,
the force imparted on the soliton cancels on the average but,
nevertheless, the stochastic nature of the force introduces a diffusive
motion of the soliton. For a moving soliton, the average force
imparted by the thermal particles does not vanish and gives rise to
a frictional force on the soliton. To include the dissipative
effects in the soliton's motion we write down the kinetic equation
for the phase space distribution function of the
soliton~\cite{landau81:kinetics}:
\begin{equation}
\frac{\partial f}{\partial t} - \frac{\partial}{\partial
p}\left(\frac{\partial H}{\partial q} f\right) + \frac{\partial}{\partial
q}\left(\frac{\partial H}{\partial p} f\right) = I_{\rm coll} ,
\end{equation}
where, for small momentum transfer, the collision integral $I_{\rm coll}$ can
be written as
\begin{equation}
I_{\rm coll} = \frac{\partial}{\partial p}\left[ Af + \frac{\partial}{\partial
p}(Bf)\right].
\end{equation}
The terms $A$ and $B$ gives rise to friction and diffusion of the
soliton respectively.
The frictional force $A$ can be computed from the following expression,
\begin{equation}
A = \int\frac{dk}{2 \pi}(-2 \hbar k) R(k)
\Big|\frac{\partial \epsilon(k)}{\hbar \partial
k}\Big| N(E,k_B T)
\end{equation}
where $N(E,k_B T)$ describes the distribution of thermal particles in the
frame of the moving soliton with velocity $v$ and the energy $E$ takes
the
value $E(k) = (\hbar k - m v)^2/2m$.
In each collision, the particle with momentum $k$ has a probability $R$
to reflect back and transfer the momentum $-2\hbar k$ to the soliton. This
momentum transfer multiplied with the number of particles coming from
each direction per unit time gives rise to a frictional force.
When the soliton is at rest, the momentum transfer on each
direction cancels on the average and as a result the friction
vanishes.

\begin{figure}[htb]
\begin{center}
\epsfig{file=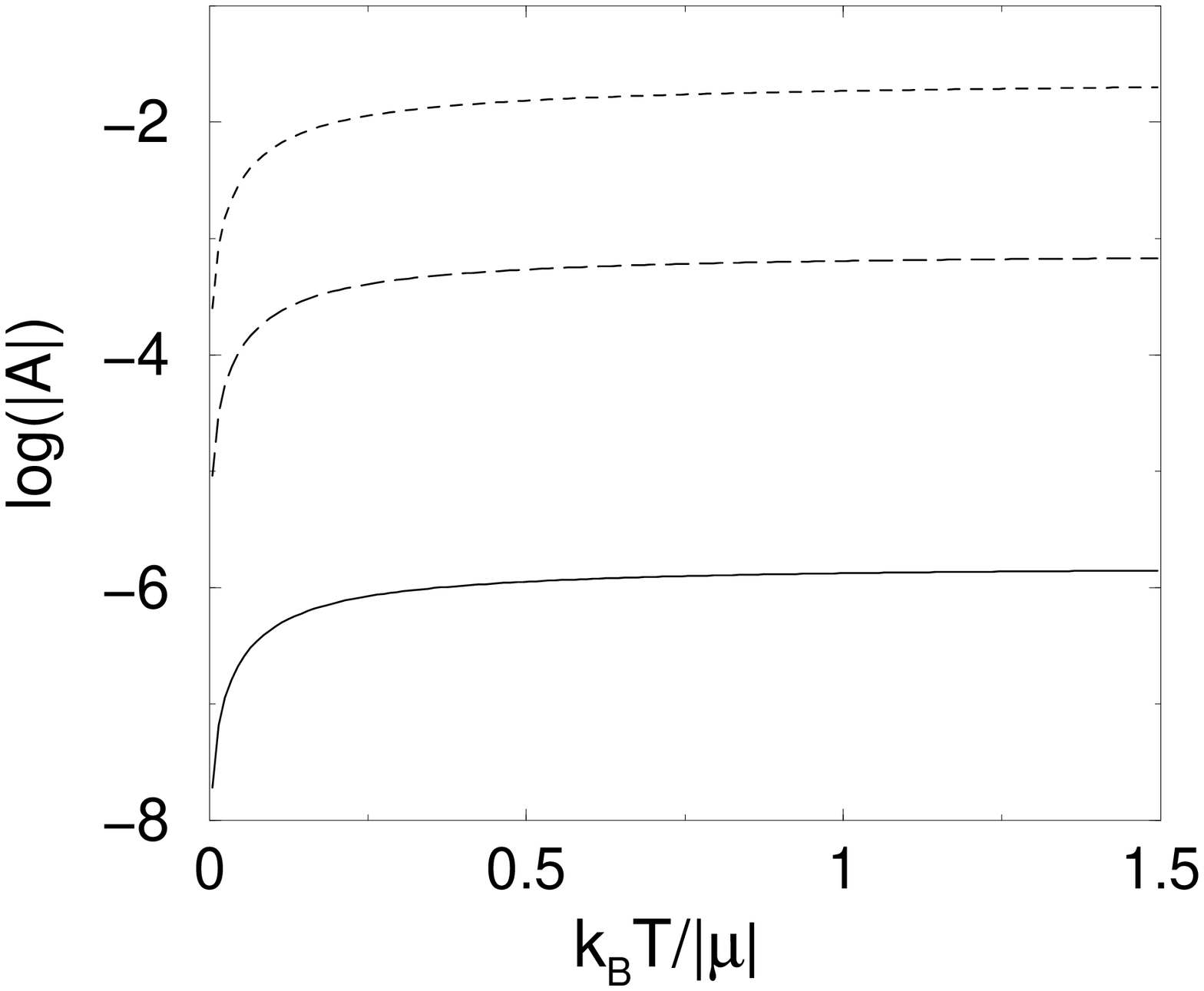,width=4cm}
\epsfig{file=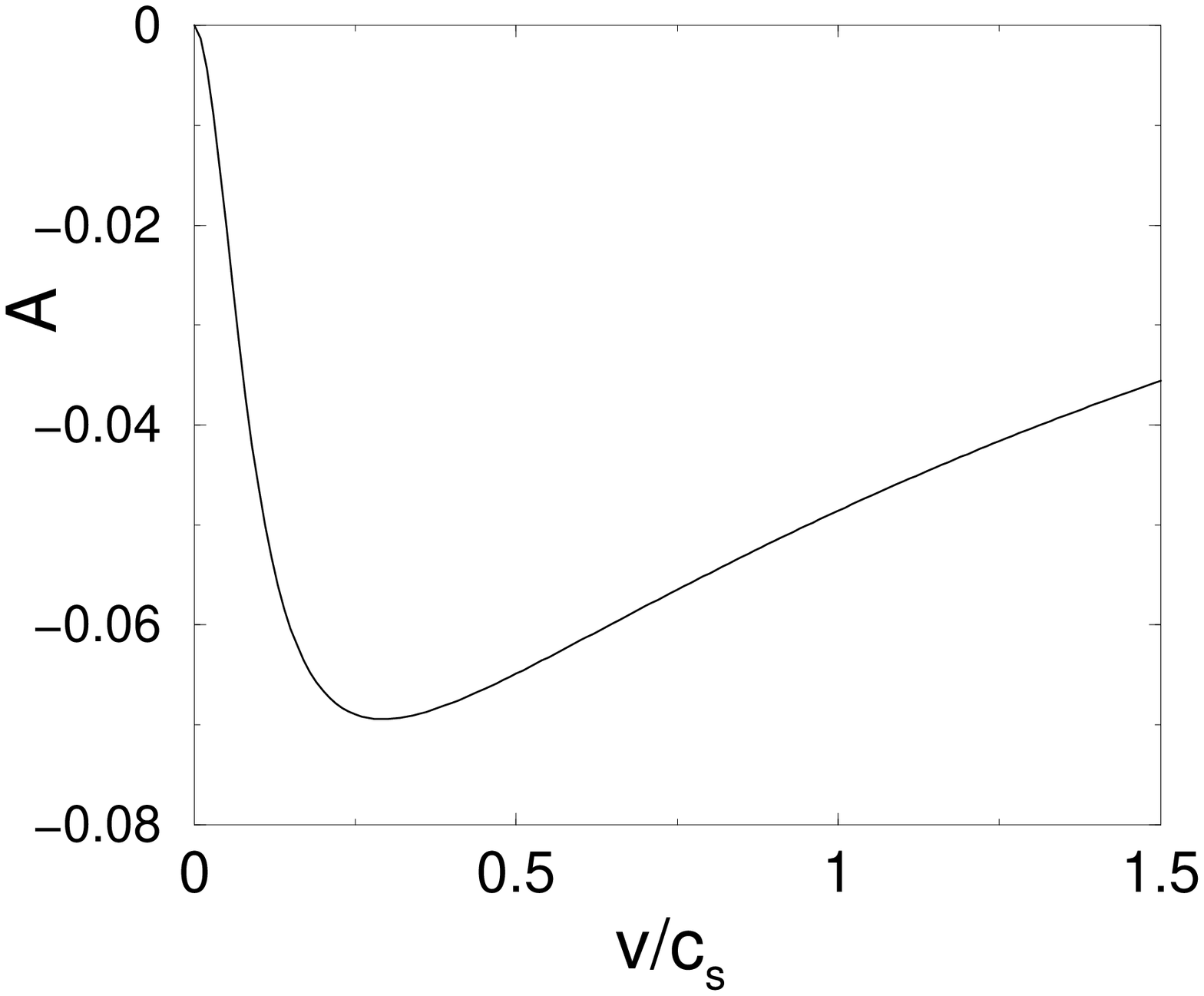,width=4cm}
\end{center}
\hskip -2cm
\vspace*{-1.2cm}
\caption{a) Logarithm of the friction force A in units of $\hbar
\omega/l$
as a function of
temperature for soliton velocity $V = 0.1\hbar/m b$, density of thermal
gas $n b = 200$, and soliton parameter
$N|a|/l =$ 0.2 (solid line), 0.3 (dashed line), 0.4 (dotted line).
b) Friction force as a function of
velocity, for $N|a|/l = 0.3$, $k_B T/|\mu| =
0.5$ (units and parameters as Fig.~2a).
\label{fig:2}}
\end{figure}

At finite temperatures the thermal atoms are distributed according to
the rules of
quantum statistics. Although thermal
equilibrium may be reached in an external trap
\cite{dunjko03ep:solitonThermo}, the subtle conditions of equilibrium
are not necessarily fulfilled in a dynamical experimental situation.
Here, we consider the motion of a soliton relative to a significantly
warmer thermal cloud of atoms. We thus can assume
a classical Boltzmann distribution of thermal atoms, $N(E,k_B T) \sim
\exp(-E/k_B T)$.
Dissipative effects of the soliton can be enhanced
by increasing the density or the temperature of the thermal cloud. We
consider the situation
where $10^4$ thermal particles are confined within a length $L = 50 b$
($\approx 70 \mu {\rm m}$ for ENS soliton with $b = 1.4 \mu {\rm m}$),
with a density of the thermal gas of $n b = 200$ ($\approx
10^{12}/{\rm cm}^3$) and the velocity
distribution of the thermal particles being controlled by changing the
temperature.
Within a certain range of the soliton velocity the
frictional force
increases linearly with velocity as seen in Fig.~\ref{fig:2}. When the
velocity is increased further,
nonlinear effects take over and the force decreases.

\begin{figure}[htb]
\begin{center}
\epsfig{file=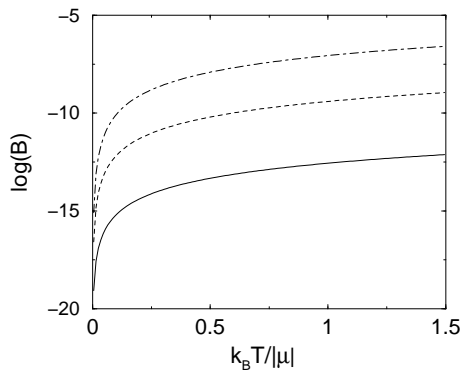,
height=5cm,width=6cm}
\end{center}
\hskip -2cm
\vspace*{-1.2cm}
\caption{Logarithm of the diffusion coefficient $B$ (in units of $\hbar m
\omega^2$) of a
soliton with zero average velocity as a
function of temperature for the soliton parameters $N|a|/l =$ 0.2 (solid
line), 0.3 (dashed line), 0.4 (dot-dashed line) and a density of the thermal
gas of $n b = 200$.
\label{fig:4}}
\end{figure}

Now we can calculate the diffusion parameter of the transport equation:
\begin{equation} \label{diffusion}
B = \int  \frac{dk}{2 \pi}2(\hbar k)^{2}R(k) \Big|\frac{\partial
\epsilon(k)}{\hbar \partial k}\Big| N(E,k_B T).
\end{equation}
This term describes the velocity fluctuations of the soliton and
gives rise to a diffusion in the momentum space. A graph is shown in
Fig.~\ref{fig:4}.

So far we considered only the low-energy elastic scattering of thermal
quasiparticles on the soliton. To restrict our discussion to the
quasi-1D case, we neglected the higher energy radial excitations $\sim
\hbar \omega$. As an additional effect, the soliton can
radiate particles if the colliding quasiparticle has higher energy
than the binding energy $|\mu|$.  Also nonlinear
collective motion of the thermal cloud and the soliton is possible
\cite{buljan05ep}. However, the elastic scattering
process discussed in this work will
dominate if the condition $k_B T <|\mu| < \hbar \omega$ is fulfilled. A
tight radial trapping potential is suitable to avoid inelastic
scattering processes.  In the ENS experiment \cite{Khaykovich2002a},
the oscillator length of radial confinement was
$l = 1.4 \mu {\rm m}$. For a soliton parameter $N|a|/l = 0.4$,
the sound velocity at the center of the soliton becomes $c_{\rm s}
\approx 2.5 {\rm mm}/{\rm s}$. If a soliton with $N \sim 10^3$
particles moves with a velocity $0.1 c_{\rm s}$, then it decelerates
$5.01 {\rm mm}/{\rm s}^2$ due to the frictional force. Finally it
stops after $0.05 {\rm s}$, travelling a distance of $6 \mu {\rm
m}$. The slowing down of a bright soliton can be observed
experimentally by suitably manipulating the density and temperature of
the thermal cloud.

Due to the friction force, the momentum $\bar{p}$ of the soliton changes
as ${d \bar{p}}/{d t} = - A(v)$. For small velocities, $A = \gamma v$
and the moving soliton stops after a time scale $\tau = mN/\gamma$.
Due to the diffusion process the energy of a resting
soliton changes as $E = (B/2 \gamma)[1 -e^{-2\gamma t/m N}]$.
For $t \ll\tau$, the energy of the soliton increases and finally it
reaches a steady state with energy
$E = B/2 \gamma$.

In conclusion, we have investigated the effects of a thermal
environment on the dynamics of bright matter-wave solitons and have
calculated the frictional force and diffusion coefficient in a
microscopic approach. Friction and diffusion effects occur due to the
deviation from the quasi-one-dimensional limit. Both of them are generally
small and can be controlled by the parameters of the system if
unattenuated propagation of solitons is desired. However, the
parameters can be chosen such that the dissipative effects become
accessible to experimental observation with currently available
techniques.

We
acknowledge enlightening
discussions with G.~Shlyapnikov who
suggested this problem.


\begin{thebibliography}{10}

\bibitem{gerton00}
J.~M. Gerton, D. Strekalov, I. Prodan, and R.~G. Hulet, Nature {\bf 408},  692
  (2000).

\bibitem{Strecker2002a}
K.~E. Strecker, G.~B. Partridge, A.~G. Truscott, and R.~G. Hulet, Nature {\bf
  417},  150  (2002).

\bibitem{Khaykovich2002a}
L. Khaykovich {\it et~al.}, Science {\bf 296},  1290  (2002).

\bibitem{hasegawa02}
A. Hasegawa, Rep. Prog. Phys. {\bf 65},  999  (2002).

\bibitem{Muryshev2002a}
A.~E. Muryshev {\it et~al.}, Phys. Rev. Lett. {\bf 89},  110401  (2002).

\bibitem{kivshar03:book}
Y. Kivshar, {\em Optical Solitons} (Acad. Press, London, 2003).

\bibitem{Wynveen2000a}
A. Wynveen {\it et~al.}, Phys. Rev. A {\bf 62},  023602  (2000);
U.~V. Poulsen and K. M{\o}lmer, {\it ibid.} {\bf 67},  013610  (2003);
J. Brand, I. H{\"a}ring, and J.-M. Rost, Phys. Rev. Lett. {\bf 91},  070403
  (2003).

\bibitem{flach:084101}
S. Flach, A.~E. Miroshnichenko, V. Fleurov, and M.~V. Fistul, Phys. Rev. Lett.
  {\bf 90},  084101  (2003).

\bibitem{Berge2000a}
L. Berg{\'e}, T.~J. Alexander, and Y.~S. Kivshar, Phys. Rev. A {\bf 62},
  023607  (2000).

\bibitem{weinstein83}
M.~I. Weinstein, Comm. Math. Phys. {\bf 87},  567  (1983).

\bibitem{carr:040401}
L.~D. Carr and J. Brand, Phys. Rev. Lett. {\bf 92},  040401  (2004);
Phys. Rev. A {\bf 70},  033607  (2004).


\bibitem{Olshanii1998a}
M. Olshanii, Phys. Rev. Lett. {\bf 81},  938  (1998).

\bibitem{cherny:043622}
A.~Y. Cherny and J. Brand, Phys. Rev. A {\bf 70},  043622  (2004).

\bibitem{carr02}
L.~D. Carr and Y. Castin, Phys. Rev. A {\bf 66},  063602  (2002).

\bibitem{mcguire:622}
J.~B. McGuire, J. Math. Phys. {\bf 5},  622  (1964).


\bibitem{landau81:kinetics}
E.~M. Lifshitz and L.~P. Pitaevskii, {\em Physical kinetics} (Pergamon, Oxford,
  1981).

\bibitem{dunjko03ep:solitonThermo}
V. Dunjko, C.~P. Herzog, Y. Castin, and M. Olshanii, cond-mat/0306514.

\bibitem{buljan05ep}
H. Buljan, M. Segev, and A. Vardi, cond-mat/0504224.

\end{thebibliography}
\end{document}